\documentclass[twocolumn,aps,prb,final,amsfonts,nobibnotes,amssymb,amsmath,floatfix,showpacs,superscriptaddress]{revtex4}

\usepackage{graphicx}

\begin{document}

%\preprint{}

\title{Probing electrodynamical properties of the edge states in a quantum Hall system by surface photovoltage spectroscopy}

\author{B. Karmakar}
\email{karmakar@tifr.res.in}
\affiliation{Tata Institute of fundamental Research, Mumbai, India}
\author{G. H. Dohler}
\affiliation{University of Erlangen, Erlangen, Germany}
\affiliation{Tata Institute of fundamental Research, Mumbai, India}
\author{B. M. Arora}
\affiliation{Tata Institute of fundamental Research, Mumbai, India}


%\author{}
%\thanks{}
%\affiliation{}


\date{\today}

\begin{abstract}
An importent question regarding the dissipation-less current carried by the edge states in a quantum Hall system is understanding the results of the electrodynamical interaction among the mobile electrons in the quantum mechanical limit under a magnetic field $B$. The interaction affects the transport parameters, the transverse electric field and the electron velocity. We have developed a new surface photovoltage spectroscopic technique to measure the parameters from the transition energies between the electron and heavy hole edge states. We observe that the measured electron velocity and transverse (Hall) electric field increase as $B^{1/2}$ and $B^{3/2}$ respectively.
\end{abstract}


\maketitle

The dispersive edge states carry dissipation-less diamagnetic current \cite{Halperin, Buttiker} along the boundary of a two-dimensional electron systen (2DES) under a perpendicular magnetic field $B$. The current is carried one-dimensionally by the filled edge states in a sub-micron conducting-strip \cite{Lorke}, which lies next to a sub-micron depletion-strip \cite{Choi, Arai} of empty edge states at the boundary. Using single particle picture, Halperin \cite{Halperin} showed the formation of current carrying edge states in the quantum Hall (QH) system. But in a real system, interacting mobile (chiral motion) electrons are present in the conducting-strip. The mobile electrons generate Hall field, which modify the transverse confining (electrostatic) electric field \cite{Guven} and electro-chemical potential \cite{Thouless} at the boundary of the 2DES. The modified transverse electric field determines the velocity of electrons and the energies of the edge states. Therefore, in the presence of a strong magnetic field, the electrodynamically interacting mobile electrons in the edge states modify the physical parameters pertaining to the dissipation-less current. However, the $B$-dependences of the physical parameters viz. the electric field, the velocity of electrons and the energies \cite{Kang} of the edge states have neither been measured nor theoretically predicted. We have developed a surface photovoltage (SPV) spectroscopic \cite{Datta} technique with a new variation so as to probe the edge states distinctively from the Landau levels (LLs) in the interior of a QH system. Our SPV technique is naturally selective to measure the average transition energies between the electron and heavy hole edge states at the interface of conducting and depletion strips. From the transition energies, we explore the electric field and the velocity of the electrons in the edge states at the interface.

\begin{figure}
\begin{center}
\includegraphics*[width=8cm]{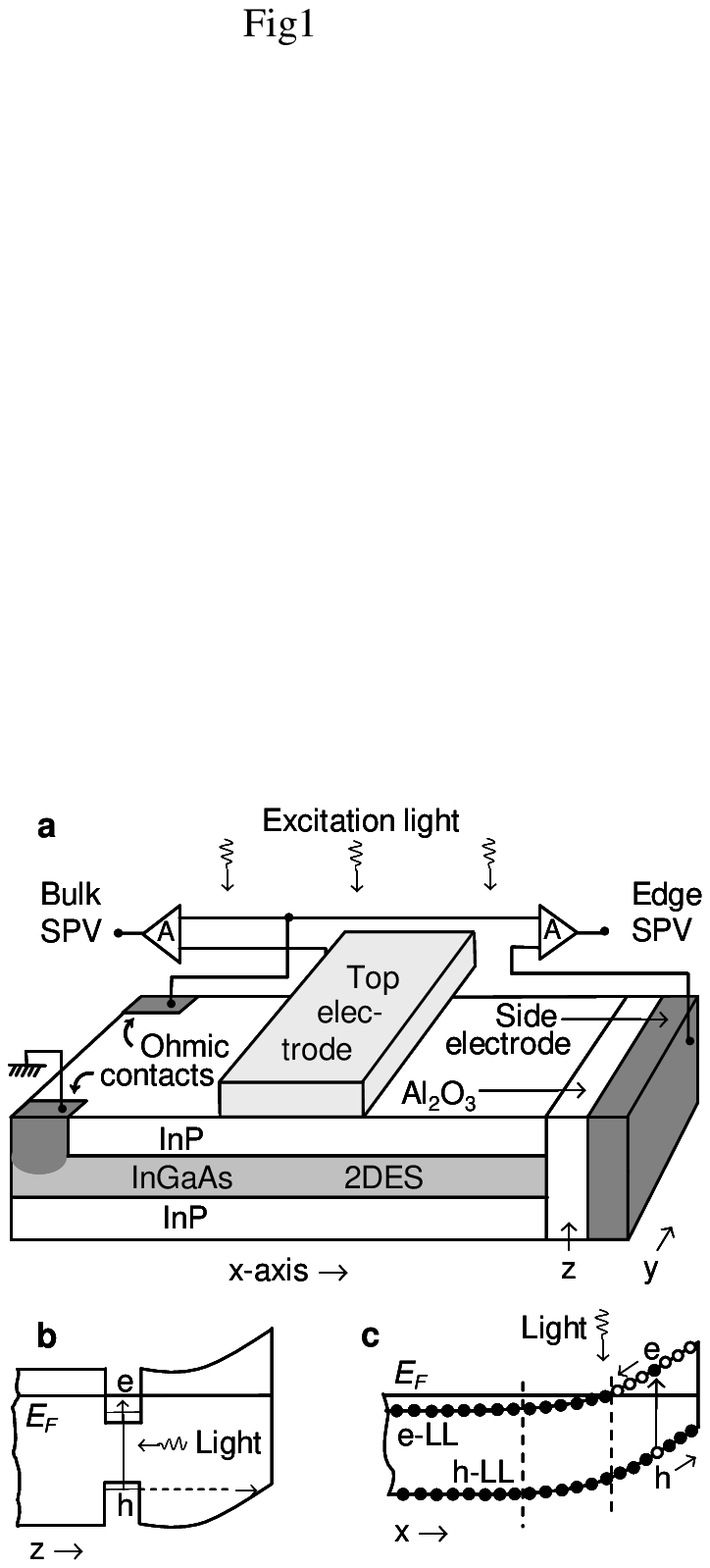}
\end{center}
\caption{$\bf{a}$) Schematic device structure and measurement setup. 'A' is unity gain buffer-amplifier circuit. $\bf{b}$) Schematic band diagram along the growth direction ($z$) to explain B-SPV signal generation. $\bf{c}$) Schematic band diagram along the plane of the QW ($x$) towards the boundary to explain E-SPV signal generation. Electron and heavy hole LLs are labeled e-LL and h-LL respectively. Vertical dashed lines separate bulk-region, conducting-strip and depletion-strip successively.}
\end{figure}

To perform the SPV experiments, in the narrow available range of energies ($763 - 816$ meV) from tunable diode laser sourses, a suitable $\textnormal{InP/In}_{0.65}\textnormal{Ga}_{0.35}\textnormal{As/InP}$ modulation doped quantum well (QW) sample is designed (Fig. 1a). The electron gas in the 2DES resides $400$ \AA~below the top surface in the $90$~\AA~thick InGaAs QW, has an electron density $n_{s} \approx 6.6 \times 10^{11}~\textnormal{cm}^{-2}$ and a mobility $\mu \approx 10^{5}~\textnormal{cm}^{2}\textnormal{/V-s}$. The sample is excited from top (Fig. 1a) with un-polarized infrared light, chopped at 20 Hz carried with an optical fiber. Electron-hole pairs are generated by electronic transitions from the filled valence band to the empty conduction band states above the Fermi energy $E_{F}$ in the QW (Fig. 1b). The bulk SPV (B-SPV) signal \cite{Datta} is generated by the tunnelling  of holes from the QW to the top surface (Fig. 1b) and is measured with excitation power density $\sim 1~\mu \textnormal{W/cm}^{2}$ (after attenuation) between an ohmic contact and the top indium-tin-oxide coated glass electrode (Fig. 1a) having transmitivity of infrared light $\sim 10 \%$. 

The novelty of our SPV technique is in the detection of the signal from the edge of the QW between an ohmic contact and a side electrode which is made of $25$~\AA~$\textnormal{Al}_{2}\textnormal{O}_{3}$ (prevent leakage) and $1000$~\AA~gold layers on the cleaved surface (Fig. 1a). In the edge SPV (E-SPV) spectroscopy, electron-hole pairs are generated by transitions from the filled valence to the empty conduction band edge states in the depletion-strip (Fig. 1c). The experiments are done in a range of energies for which the electron LLs are empty at the edge and available for optical transition, but are filled in the interior, forbiding optical transition. As a result, the photo-absorption between the quantized states in the depeltion-strip occur selectively. After photo-absorption, generated electron-hole pairs can either recombine by emitting photon or relax to their minimum position (Fig. 1c). In our experiment, the second process gives rise to E-SPV signal. The generated E-SPV signal is $\sim 100~ \mu \textnormal{V}$ at excitation power density $\sim 30~\textnormal{nW/cm}^{2}$. Very sensitive unity gain buffer amplifier circuit (Fig. 1a) made of electrometer grade operational amplifier is used to measure the SPV signals. The shield of the input cable is driven by the output of the amplifier to nullify the input cable capacitance, which enable the circuit to measure photovoltage generated from sub femto-coulomb charge. Differential voltage is measured between two such amplifiers, one connected to ohmic contact and the other to side (for E-SPV) or top electrode (for B-SPV).

\begin{figure}
\begin{center}
\includegraphics[width=7cm]{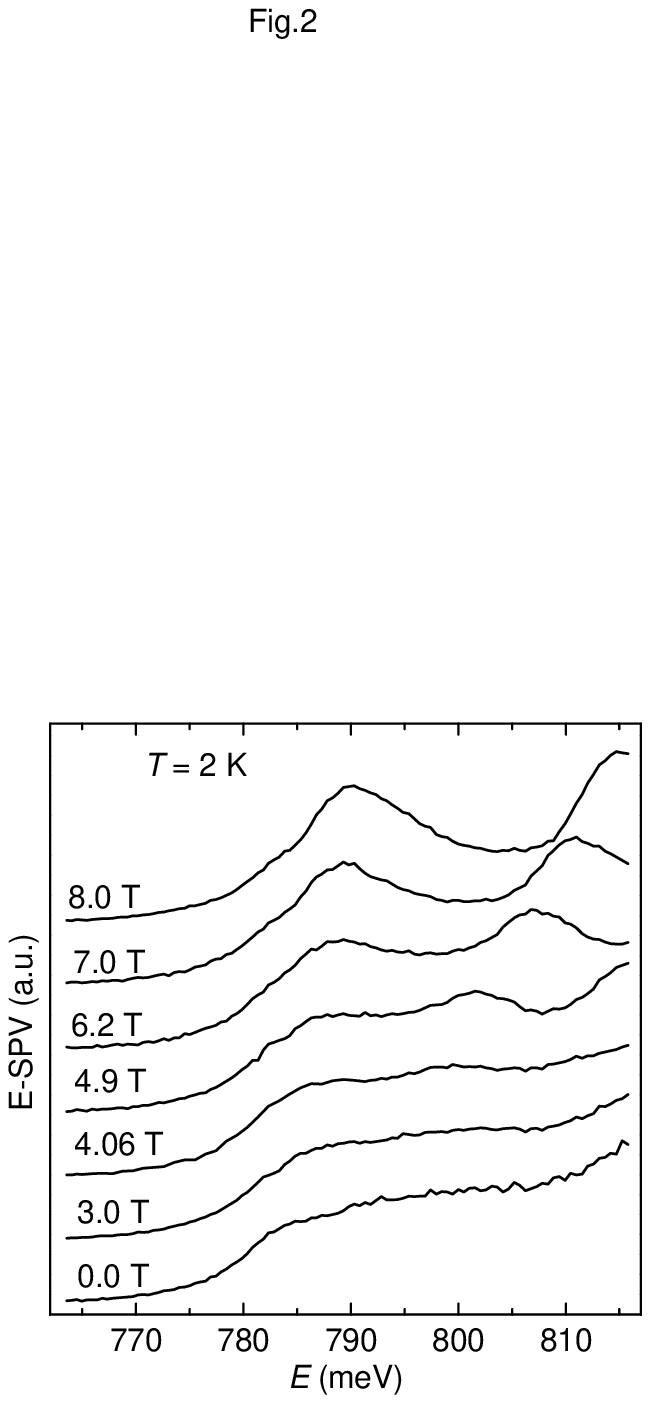}
\end{center}
\caption{Plots of the E-SPV spectra at differnt $B$ fields (measured without passing any external current). The plots are shifted vertically for clarity.}
\end{figure}

The E-SPV spectra at selected $B$ values are plotted in Fig. 2. The distinct peaks in the E-SPV spectra are progressively resolved and shift to higher energy with increasing $B$. However, Zeeman splitting is not resolved indicating that the Zeeman energy is less than the broadening of the peaks. 

The $B$ values for E-SPV experiments are selected from two-terminal magneto resistance (2TMR) plot (Fig. 3a), identified with the filling fractions $\nu$, where $\nu = n_{s}/(eB/h)$ is defined as the number of filled LLs and $eB/h$ is the density of states (DOS) of a spin-split LL. The 2TMR is measured between two ohmic contacts (Fig. 1a) and it is a combination of Hall $(R_{H})$ and longitudinal $(R_{L})$ resistances. At the 2TMR plateaus, $R_{L}= 0$, such that the measured resistance is purely Hall resistance of well defined values $h/\nu e^{2}$  with integer filling fraction $\nu$. At the plateau-to-plateau transitions, $R_{L}$ goes through a maximum and shows a hump at the beginning of each plateau.

The B-SPV experimant at finite $B$ is discussed in appendix A. The E-SPV and B-SPV spectra at $B = 0$ are compared in Fig. 3b. Blue-shift of the B-SPV spectrum, compared to the E-SPV spectrum, results from the band filling of electrons \cite{Davies}. This comparison proves that the edge states are probed distinctively from the states in the interior. The E-SPV spectrum for $B = 0$ rises at an energy characteristic of the band gap of the QW and is flat at higher energies. The SPV signal results from generation of electron-hole paires by photo-absorption and charge separation. Because of low measurement temperature, excitons do not break into electron-hole pairs \cite{Tari}. Therefore, the E-SPV spectrum ($B = 0$) is free from excitonic peak like feature (Fig. 3b). Also, we do not see Franz-Keldysh oscillation in the E-SPV spectrum because of low electric field at the edge. Moreover, the observed E-SPV ($B = 0$) spectrum is less sharper (Figs. 3b\&c) than an ideal sample, possibly due to disorder and the edge electric field. In order to determine the band gap energy $E_{g}$, the E-SPV spectrum is numerically smoothed and numerical derivative of the smoothed curve is taken (Fig. 3c). The maximum of the derivative is well defined, which gives the band gap energy $E_{g} = 780.9 \pm 0.3$ meV at temperature 2 K.

\begin{figure}
\begin{center}
\includegraphics[width=8cm]{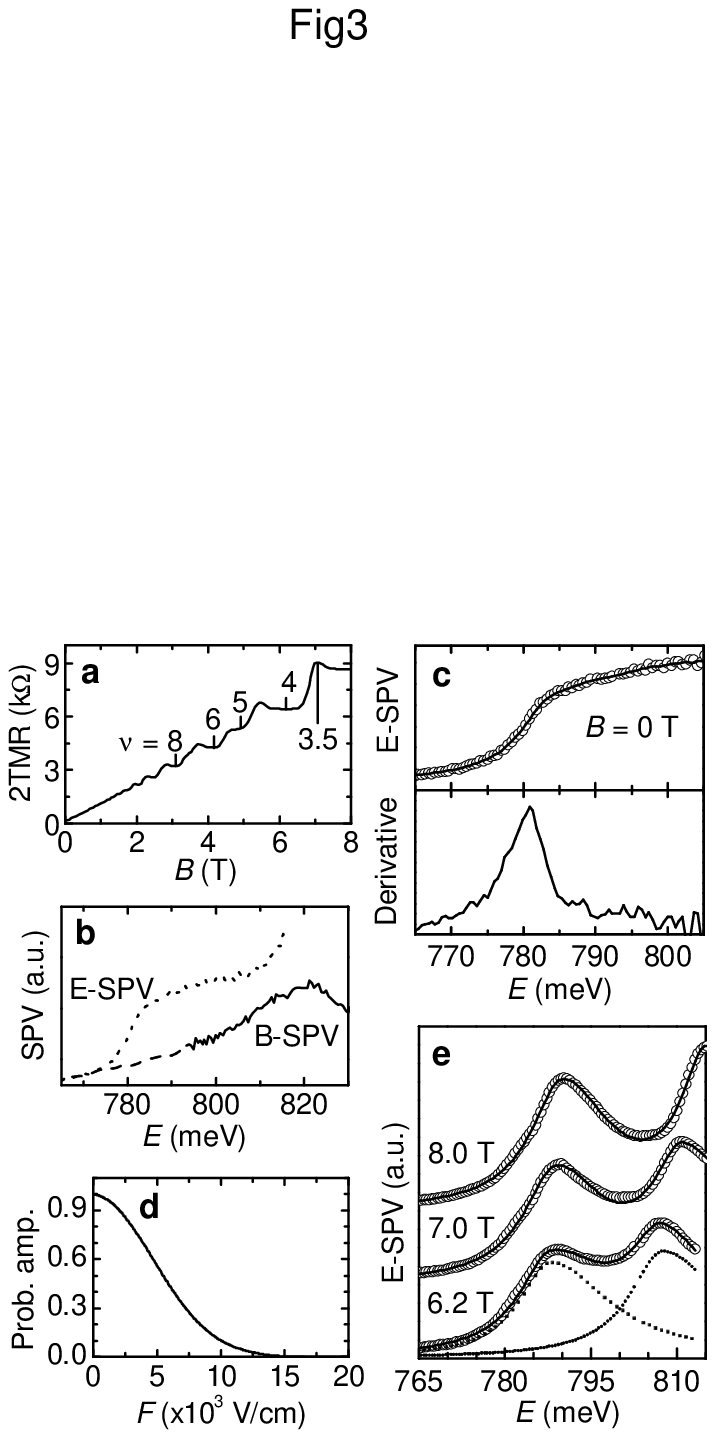}
\end{center}
\caption{$\bf{a}$) Plot of 2TMR at 1.8 K, $\nu$ is filling fraction. $\bf{b}$) Comparison of E-SPV and B-SPV at $B = 0$. At the higher energies (792-830 meV), the B-SPV signal is measured by conventional method \cite{Datta} at 8 K. $\bf{c}$) Upper plot is numarically smoothed E-SPV spectrum (solid line), circles represent data points. Lower plot is derivative of the numarically smoothed spectrum. $\bf{d}$) Plot of normalized transition probability amplitude at $B = 8$ T with $F$ considering $m_{e}^{*} \approx 0.047m_{e}$, $m_{hh}^{*} \approx   0.173m_{e}$ and localized wavefunction along $x$. $\bf{e}$) Curves fitted to the E-SPV spectra. Circles represent the numerically smoothed E-SPV spectra and the solid lines are the fitted curves. Dotted lines are two components of the fitted curve for $B = 6.2$ T spectrum.}
\end{figure}

The energetics of the edge states from edge spectra (Fig. 2) can be understood in terms of single particle picture in the depletion-strip because of the absence of the electron-electron interaction. Considering constant electric field $F$ along $x$ towards the boundary (Fig. 1a) in the depletion region, single particle Hamiltonian of an electron without the Zeeman energy can be written as 

\begin{equation}
H =\frac{1}{2m_{e}^{*}}(\mathbf{p}-e\mathbf{A})^{2}+eFx
\end{equation}
with the Landau gauge of the magnetic vector potential $\mathbf{A}(0,xB,0)$, where $m_{e}^{*}$ is the electron effective mass and $\mathbf{p}$ is the momentum vector. Using Eq.1 the energy $E_{n,k_{y}}$ for nth LL is deduced \cite{Davies} as

\begin{equation}
E_{n,k_{y}} = \left(n + \frac{1}{2}\right) \hbar \omega_{c} - \hbar\left(\frac{F}{B}\right)k_{y} - \frac{1}{2}m_{e}^{*}\left(\frac{F}{B}\right)^{2}
\end{equation}
where  $\omega_{c} = eB/m_{e}^{*}$ is the cyclotron frequency, $k_{y}$ is the quantized wave vector along $y$ (boundary). The second and third terms in Eq.2 are respectively linear dispersion and the energy correction introduced by the field $F$. Solving the Schrödinger equation for heavy hole and using Eq.2, the edge transition energy $E_{Tn}$ from nth hole LL to nth electron LL is deduced as

\begin{equation}
E_{Tn} = E_{g} + \left(n + \frac{1}{2}\right)\hbar \omega_{r} - \frac{1}{2}(m_{e}^{*} + m_{hh}^{*})\left(\frac{F}{B}\right)^{2}
\end{equation}
where $\omega_{r} = eB/m_{r}$ is reduced cyclotron frequency, $m_{r} = m_{e}^{*}m_{hh}^{*}/(m_{e}^{*}+m_{hh}^{*})$ is reduced mass and $m_{hh}^{*}$ is heavy hole mass. Momentum conservation eliminates the dispersion term in Eq.3. The third term in Eq.3, independent of $n$, is the electric field (EF) correction to the energy $E_{Tn}$. If a weak parabolic potential in the depletion-strip is considered, the edge transition energy remains same as in Eq.3, only $F$ increases linearly with $x$.

The variation of $F$ with $x$ in the depletion-strip of a real sample is not simple. Away from the interface of conducting and depletion strips (along $x$, Fig. 1c) $F$ becomes larger due to the parabolic and the higher order terms (Taylor expansion) of the edge potential. Therefore, in the depletion-strip the value of $F$ is minimum at the interface of conducting strip and maximum at the boundary of the 2DES. The variation of $F$ in the depletion-strip introduces variation of optical transition probability along $x$. Due to the field $F$, the total shift between the centre of masses \cite{Davies} of electron and hole, having same wave vector $k_{y}$, is $(m_{e}^{*} + m_{hh}^{*})F/eB^{2}$. The transition probability includes the overlap integral of electron-hole wave functions and is plotted for a constant field $B = 8$ T with respect to $F$ in Fig. 3d. Since significantly larger fields $(F)$ are present in the depletion-strip near the sample boundary, this region would hardly contribute to the transition between the valence to conduction band edge states according to Fig. 3d. Therefore, we measure the transition energies, in the first approximation, at the minimum electric field region in the depletion-strip, which lies just at the boundary of the conducting-strip. Thus, E-SPV experiment selectively measures the average transition energy near the interface of conducting and depletion strip. The line shape of the E-SPV spectra are assymetric because of the variation of transition probability with $F$ in the depletion-strip.

Asymmetric Lorentzian curves are fitted to the numerically smoothed E-SPV spectra (Fig. 3e) and the peak positions of the fitted curves give the average edge transition energies $E_{T1}$ and $E_{T0}$ near the interface of conducting and depletion strip. The edge transition energies are plotted in Fig. 4 with $B$ and we clearly see a linear dependence of $E_{T1}$ and $E_{T0}$ on $B$ in the QH regime. More importantly, we verify experimentally the inequality $1/3(E_{T1}-E_{g}) > (E_{T0}-E_{g})$ for a given $B$, which proves the existence of the EF correction (Eq.3). The observed inequality can not result from band non-parabolicity, which would yield the inequality $1/3(E_{T1}-E_{g}) < (E_{T0}-E_{g})$ as the cyclotron energy would decrease with increasing energy due to the increasing effective mass $m_{e}^{*}$.

\begin{figure}
\begin{center}
\includegraphics[width=7cm]{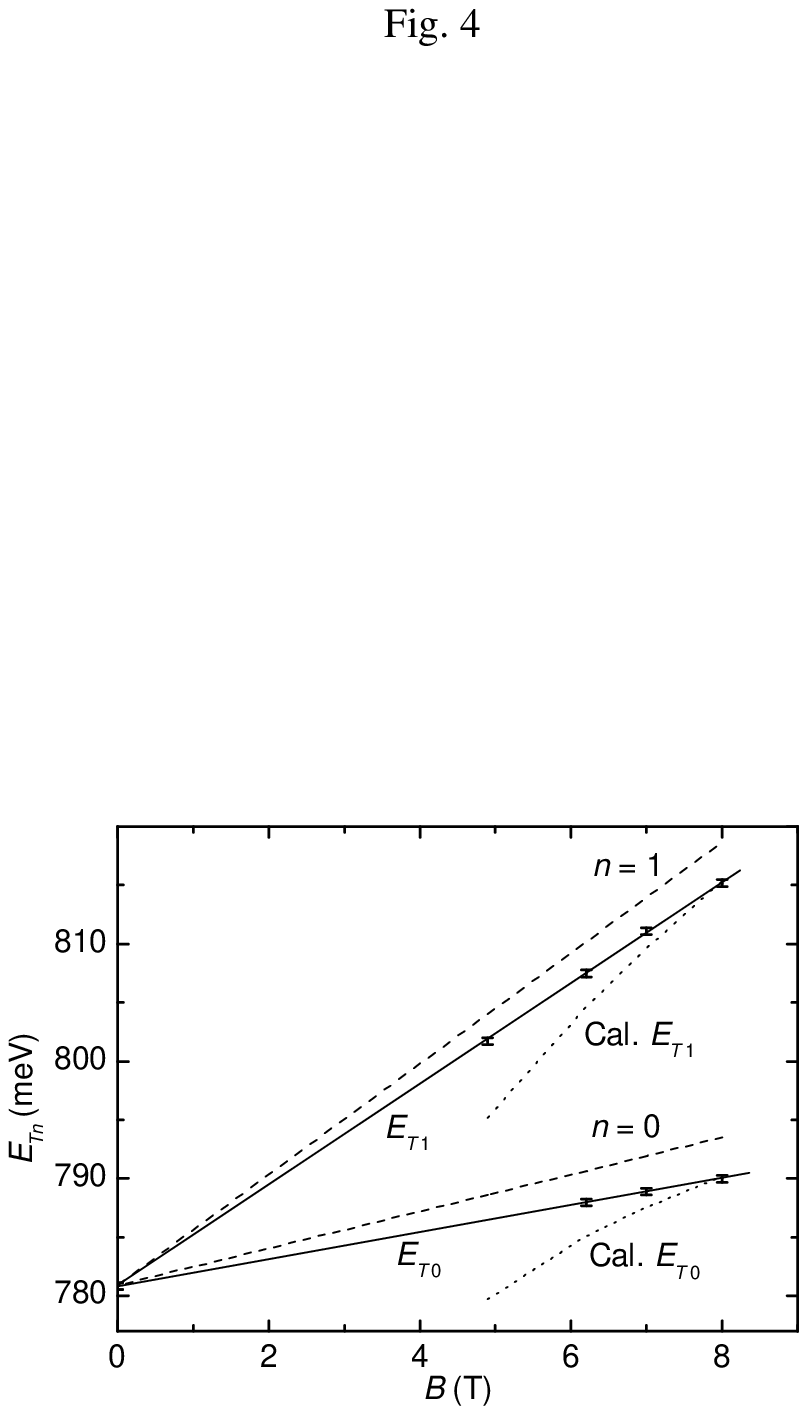}
\end{center}
\caption{Plots of the energies $E_{T1}$ and $E_{T0}$ with $B$. Dashed lines are the bulk transition energies for $n = 0$ and $n = 1$ without EF correction. Dotted lines are the calculated curves for $E_{T1}$ and $E_{T0}$ (Cal. $E_{T1}$ and $E_{T0}$) considering fixed $F = 5.84 \times 10^{3}$ V/cm in Eq.3.}
\end{figure}

The linear $E_{Tn}$ plots passing through $E_{g}$ at $B = 0$ is a key finding. The linearity of $E_{Tn}$ in the QH regime implies that the EF correction is either constant or proportional to $B$. The $E_{Tn}$ plots passing through $E_{g}$ at $B = 0$ confirms the linear variation of the EF correction with $B$ (Fig. 4). Using Eq.3 and the slopes of the $E_{Tn}$ plots, we obtain cyclotron energy per unit $B$ field as $E_{r} = \hbar \omega_{r}/B = 3.15 \pm 0.06~\textnormal{meV/T}$ and find $m_{r} = (0.037 \pm 0.001)m_{e}$. Furthermore, we obtain the EF correction per unit $B$ field as $E_{ef}  = 0.42 \pm 0.06~\textnormal{meV/T}$.

From the linearity of the EF correction, we get the most fundamental finding that the average electric field $F_{av}$ at the interface of conducting and depletion strip increases as a power law $B^{3/2}$. The striking consequence of the finding is that the depletion width $l_{d}$ shrinks with increasing $B$, since $l_{d} \sim \phi_{b}/F_{av}$ and the surface barrier potential $\phi_{b}$ ($\phi_{b} \approx 250$ meV for InGaAs Ref.\cite{Bhattacharya}) remains constant by the pinning of $E_{F}$. From the EF correction, we get an estimate for $F_{av} \approx 5.8 \times 10^{3}~\textnormal{V/cm}$ at $B = 8~\textnormal{T}$, using $m_{e}^{*} \approx 0.047m_{e}$ Ref.[{\cite{Sugawara}] and $m_{hh}^{*} \approx   0.173m_{e}$.

For a clear understanding about the EF correction, we plot the bulk transition energies (without the EF correction) $E_{g} + 1.5 \hbar \omega_{r}$ for $n = 1$ and $E_{g} + 0.5 \hbar \omega_{r}$ for $n = 0$ in Fig. 4. The red shift of the edge transition energies $E_{T1}$ and $E_{T0}$ due to the EF correction is clearly seen. Furthermore, we plot the calculated curves of $E_{T1}$ and $E_{T0}$ in Fig. 4 using Eq.3 and keeping a constant electric field (hard potential) $F_{av} = 5.84 \times 10^{3}~\textnormal{V/cm}$ deduced at $B = 8~\textnormal{T}$. The calculated curves deviate much beyond the error bar from the experimental points. Therefore, the linearity in $E_{Tn}$ with $B$ can only be explained by a $B-dependent$ electric field $F_{av} \propto B^{3/2}$ (soft potential).

The dependence $F_{av} \propto B^{3/2}$ shows that the edge potential is soft i.e. the potential profile at the edge depends on charge redistribution \cite{Guven} with $B$ in the conducting-strip because of the counter balance of the local electrostatic and electromagnetic forces. Therefore, the $B$-dependence of the electric field $F_{av}$ results from a $B-dependent$ screening effect in the conducting-strip. 

Due to the screening effect, the electron density in the conducting-strip gradually decreases from the bulk value $n_{s}$ to zero \cite{Chklovskii} at the interface of conducting and depletion strips. At the interface, application of Gauss's theorem shows that the electric field is continuous. Therefore, the electric field component along $x$ in the conducting-strip near the interface is same as the measured $F_{av}$ in the depletion-strip. Hence, our finding has another consequence, viz., the average velocity of electrons $v_{av} = F_{av}/B$ near the interface increases as a power law $B^{1/2}$. The average velocity of electrons $v_{av} \approx 7.3 \times 10^{6}$ cm/s is estimated at 8 T. 

Surprisingly, the dependence $v_{av} \propto B^{1/2}$ is same as the $B$-dependence of the Larmor velocity of electrons $\omega_{c}l_{0} \sim B^{1/2}$ ($\omega_{c}l_{0} \approx 2.71 \times 10^{7}$ cm/s at $8$ T), $l_{0} = (h/eB)^{1/2}$ is the magnetic length. Single particle quantum mechanical analysis (for example Eq.2) gives the velocity of electrons as $F/B$. So far there is no theoretical analysis which relates the velocities $F/B$ and $\omega_{c}l_{0}$. In reality, the charge redistribution with $B$ at the edge is taking place in such a way that the dependence $F_{av} \sim B^{3/2}$ (hence $v_{av} \sim B^{1/2}$) emerges. But the detailed understanding of our results requires a self-consistent calculation including $B-dependent$ screening.

We have established that SPV spectroscopy is a unique technique to probe the edge states. As an extension of our experiment, using the E-SPV spectroscopy, the interplay between the Hall current (external) and the chiral current \cite{Shizuya} can be probed. Apart from that, our spectroscopic technique has numerous potential applications to probe Fermi liquid to Luttinger liquid transition \cite{Hilke} and the energetics of the Luttinger liquid \cite{Wen} in fractional QH systems. Using our spectroscopic technique the relaxation process, which generates E-SPV signal, in the edge states can also be studied. Other applications of our SPV technique are discussed in appendix B.

Our experiment displays the general electrodynamics in the quantum mechanical limit i.e. the effect of the $B-dependent$ interaction among the mobile electrons in the edge states as the power law dependences of the transverse (Hall) electric field and the electron velocity on $B$. The observed power-law dependence has to be considered in the theory of the electrostatics and reconstruction of the edge states \cite{Chklovskii} to understand the electro-chemical potential distribution,which is very important in the exact quantization of the Hall resistance in the integer and fractional QH effects.

\appendix

\section{}
The B-SPV signal is measured with varying $B$ at the excitation energy $815.8$ meV ($1520$ nm) and is plotted in Fig. 5 along with the 2TMR for comparison. The excitation energy $815.8$ meV is the closest available energy for the transition from filled valence to empty conduction band LLs at the Fermi energy in the interior. The B-SPV experiment maps the LLs as each LL crosses the Fermi energy $E_{F}$. The B-SPV signal is measured by keeping the sample edges covered with aluminium foil. At low magnetic field, we see oscillations in the B-SPV signal (inset of Fig. 5) analogous to the Shubnikov-de Haas oscillations. In the QH regime, sharp peaks appear in the B-SPV signal at even-$\nu$ to odd-$\nu$  plateau to plateau (PP) transitions corresponding to spin-split LLs. The widths and the heights of successive B-SPV peaks increase with $B$ due to increase of the density of states ($eB/h$) of the LLs. Similar B-SPV peaks at odd-$\nu$ to even-$\nu$ PP transitions are not observed because the transitions require higher energy due to higher Landau and Zeeman energies, which will be published elsewhere. In addition, sharp rise and fall of the B-SPV peaks are seen.

\begin{figure}
\begin{center}
\includegraphics[width=8cm]{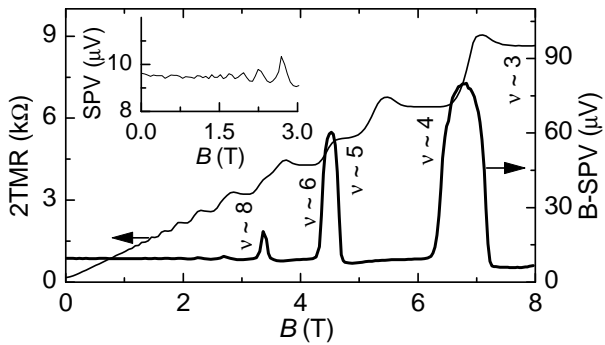}
\end{center}
\caption{Plot of B-SPV signal and 2TMR with $B$. Inset shows the B-SPV oscillations at low $B$.}
\end{figure}

\begin{figure}
\begin{center}
\includegraphics[width=7cm]{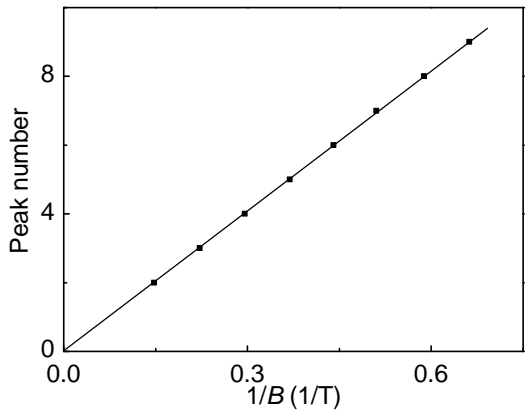}
\end{center}
\caption{Plot of B-SPV peak number versus $1/B$.}
\end{figure}

The unusual shape of the B-SPV peaks in Fig. 5 can be easily understood. In the presence of disorder, degenerate LL broadens into a band of extended states surrounded by localized tail states which are sharply separated by the mobility edges. The electron and hole localized states would be spatially separated; hence overlap integral between electron-hole wave functions is small. As a result, the probability of transition between them is also small. Moreover, the transition probability between localized states and extended states is also negligible because of small overlap integral. The sharp rise and fall of the B-SPV peaks in Fig. 5 are seen because the B-SPV experiment maps the joint-density of extended states. As a result, we see sudden onset of the optical transition between the extended states as well as sudden fall with increasing $B$ as the mobility edges cross the $E_{F}$. Therefore, in the B-SPV experiment, we can measure directly sum of the widths of electron and hole extended states. 

Successive B-SPV peaks in Fig. 5 are assigned integer numbers in increasing order from higher to lower $B$, and corresponding $B$ values of the peaks are determined. The peak number versus $1/B$ plot (Fig. 6) is linear, having a slope of $n_{s}h/2e$. The slope of this plot provides an estimate of the electron density $n_{s} = (6.60 \pm 0.03)\times 10^{11}~\textnormal{cm}^{-2}$. 

The B-SPV and E-SPV spectra at $B = 0$ are compared in Fig. 3b. The energy of transition $E_{TF} \approx 822$ meV from valence band to empty conduction band states near the $E_{F}$ is determined from the peak in the B-SPV spectrum which is shifted by $n_{s}h^{2}/m_{r}$ in comparison to $E_{g}$. Using the values of $E_{g}$, $n_{s}$ and $m_{r}$, the transition energy $E_{TF}$ is estimated as $823.6 \pm 1.7$ meV, which matches reasonably. The B-SPV spectrum in the QH regime and the consequences of the results will be publishd elsewhere.

\section{}
The B-SPV experiments are done in weakly invasive manner such that the generated B-SPV signal ($\sim 100~\mu$V) is much less than the surface barrier potential ($\sim 300$ mV for InP) with low excitation power density, $\sim 1~\mu\textnormal{W/cm}^{2}$ after attenuation in the top electrode. The B-SPV experiments require lower excitation power density ($\sim 1~\mu\textnormal{W/cm}^{2}$) compared to that used in the conventional transmission experiments \cite{Aifer} by about a factor of 1000 and in the inelastic light scattering experiments \cite{Moonsoo} by about a factor of 100. Very weak perturbation of the system enables us to measure the widths of the joint density of states directly (Fig. 5). Therefore, in the quantum Hall system the B-SPV experiment can be used to study the quantum criticallity \cite{Pruisken}, in that the nature of the dependence of the width of the extended states with temperature is studied. Using B-SPV experiment, the manybody excited states \cite{Pinczuk} above the Fermi energy in the fractional quantum Hall system can also be studied.

\begin{acknowledgments}
We greatfully acknowledge helpful discussions with D. Dhar, K. L. Narasimhan, B. Bansal and S. Bhattacharya. We thank R. Bhat for providing sample, E. L. Ivchenko for useful comments, A. P. Shah for help in device preparation, S. Ghosh and J. Bhattacharya for conventional B-SPV measurement used in Fig. 3b.
\end{acknowledgments}


\end{document}